\documentclass[authoryear,12pt]{elsarticle}
\usepackage{amssymb}
\usepackage{graphicx}
\usepackage{epstopdf}
\usepackage{slashbox}
\usepackage{picture}
\usepackage{natbib}

\journal{Computational Statistics and Data Analysis}
\begin{document}

\begin{frontmatter}
\title{Approximate Bayesian Computing for Spatial Extremes}

\author{Robert J. Erhardt, Richard L. Smith}

\address{Department of Statistics and Operations Research\\
318 Hanes Hall CB \#3260\\
University of North Carolina at Chapel Hill\\
Chapel Hill, North Carolina 27599\\
919-843-6024\\
erhardt@email.unc.edu, rls@email.unc.edu}

\begin{abstract}
Statistical analysis of max-stable processes used to model spatial extremes has been limited by the difficulty in calculating the joint likelihood function.  This precludes all standard likelihood-based approaches, including Bayesian approaches.  In this paper we present a Bayesian approach through the use of approximate Bayesian computing.  This circumvents the need for a joint likelihood function by instead relying on simulations from the (unavailable) likelihood.  This method is compared with an alternative approach based on the composite likelihood.  When estimating the spatial dependence of extremes, we demonstrate that approximate Bayesian computing can provide estimates with a lower mean square error than the composite likelihood approach, though at an appreciably higher computational cost.  We also illustrate the performance of the method with an application to US temperature data to estimate the risk of crop loss due to an unlikely freeze event.
\end{abstract}

\end{frontmatter}

\section{Introduction}
Modeling of spatial extremes is motivated by the need to model and predict environmental extreme events such as hurricanes, floods, droughts, heat waves, and other high impact events.  Though the data have a natural spatial domain, standard spatial statistics methods may fail to accurately model extremes.  Models specifically designed for extremes are better suited.  The urgency of focusing on extremes is increased when one considers the potential influence of climate change on the probability of such high impact events.  We consider point referenced data, usually taken as daily or hourly measurements $y_{t,d}$ at locations $d=1, ..., D$ for time points $t=1, ..., T$.  When modeling extremes, as a first step one takes block maxima over some temporal block (usually one year) and obtains block maxima data $y_{i,d}$ where $i$ is the block.  In the environmental setting, the data would typically be annual maxima at each of $D$ locations.

For a single location, univariate extreme value theory provides a full range of tools to analyze the data.  This theory is well developed and documented \citep{coles01, dehaan06, embrechts97, resnick87, resnick07}.  When one considers several locations at once, multivariate extreme value theory is a natural extension.  Multivariate models often work well for lower dimensions, but if the data have a natural spatial domain and the dimension grows rapidly, spatial extreme value theory becomes useful.  Spatial extremes are the infinite dimensional generalization of multivariate extremes.  The goal then is to fit these block maxima data to a spatial process model so that the spatial dependence may be estimated.  One promising class of models are max-stable processes.  These processes arise as the limiting distribution of the maxima of independent and identically distributed random fields.  A number of max-stable process models have been described \citep{schlather02,kabluchko09} and one unpublished model was described by Smith in 1990.  The statistical analysis of these models is limited by the unavailability of the joint likelihood function.  However, the bivariate distributions are available in closed-form.  This allows one to write down the pairwise log-likelihood, which is the sum (taken over all unique pairs of locations) of all bivariate log-likelihoods, and is thus also a composite log-likelihood.  Numerical maximization of the composite likelihood yields estimates of the parameters which are consistent and asymptotically normal \citep{padoan10, lindsay88}.  Maximum composite likelihood estimation has been the only method so far for analyzing max-stable processes which is widely applicable, implemented computationally (\texttt{R} package \texttt{SpatialExtremes}), and for which a viable asymptotic theory exists.

In this paper we develop a Bayesian alternative for analyzing the dependence of spatial extremes.  It circumvents the need for the joint likelihood, and instead relies only on simulations.  This approach, termed approximate Bayesian computing, has been successfully applied in many areas, including extreme values \citep{bortot07}.  We show three implementations of the approximate Bayesian computing approach for analyzing spatial extremes.  The first two rely on the bivariate distribution function, and like the composite likelihood approach they consider the spatial dependence through all unique pairs of locations.  The third and most successful approach extends beyond pairs, and is able to consider higher order k-tuples for $k \ge 3$.  This feature is an important benefit of the approximate Bayesian computing approach over all pairwise approaches.

We show that the approximate Bayesian computing method can result in a lower mean square error compared to the competing composite likelihood approach when estimating the spatial dependence.  We also discuss how this Bayesian approach naturally incorporates parameter uncertainty into predictions, which is a central task in the field of extremes.  The method is computationally intensive, but the implementation based on independent sampling can be carried out in parallel, and we demonstrate the use of a more computationally efficient adaptive implementation as well.

\section{Extremes}
\subsection{Univariate and Multivariate Extremes}
Let $Y_{1}, ..., Y_{n}$ be independent and identically distributed univariate random variables with some distribution function $F$ and let $M_{n} = \max(Y_{1}, ..., Y_{n})$ be the maximum.  If $M_{n}$ converges to a non-degenerate distribution under renormalization as
\[
pr \left(\frac{M_{n} - b_{n}}{a_{n}} \leq y \right) = F^{n}(a_{n}y + b_{n}) \rightarrow G(y) \mbox{ as } n \rightarrow \infty
\]
for some sequences $a_{n} > 0$ and $b_{n}$, then $G$ must be a member of the Generalized Extreme Value family, with distribution function
\[
G(y) = \exp \left\{-\left(1+ \xi \frac{y - \mu}{\sigma}\right)_{+}^{-1/\xi} \right\}.
\label{eq:Generalized Extreme Value}
\]
Here $a_{+} = \mbox{max}(a,0),$ and $\mu, \sigma >0, $ and $\xi$ are the location, scale, and shape parameters, respectively \citep{coles01}.  The sign of the shape parameter $\xi$ corresponds to the three classical extreme value distributions: $\xi>0$ is Fr\'{e}chet, $\xi<0$ is Weibull, and $\xi \rightarrow 0$ (in the limit) is Gumbel.  The Fr\'{e}chet case corresponds to a heavy tailed distribution, Gumbel is intermediate, and Weibull has a bounded upper limit.  The Generalized Extreme Value distribution also holds for modeling minima, since one may always write $\min(Y_{1}, ..., Y_{n}) = - \max(-Y_{1}, ..., -Y_{n}).$

One property of the Generalized Extreme Value distribution is that it is max-stable: If $Y_{1}, ..., Y_{n}$ are independent and identically distributed from $G$, then $\max(Y_{1}, ..., Y_{n})$ also has the same distribution with only a change in location and scale, as $G^{n}(y) = G(A_{n}y + B_{n})$ for constants $A_{n}$ and $B_{n}$.  A distribution is a member of the Generalized Extreme Value family if and only if it is max-stable \citep{leadbetter83}.  A special case of the Generalized Extreme Value family is the unit-Fr\'{e}chet, with distribution function $G(y) = \exp(-{1}/{y})$.  Any member of the Generalized Extreme Value family may be transformed to have unit-Fr\'{e}chet margins as follows: if $Y$ has a Generalized Extreme Value distribution, then a new variable $Z$ may be defined as
\[
Z = \left(1 + \xi\frac{Y - \mu}{\sigma}\right)_{+}^{1/\xi}
\]
and $Z$ has unit-Fr\'{e}chet margins.  If the parameters are unknown, they may first be estimated and then the transformation to $Z$ is taken.  When we model multivariate or spatial extremes, there is no loss in generality when one assumes the margins are all unit-Fr\'{e}chet.  In practice, one would first estimate all marginal distributions and transform to unit-Fr\'{e}chet, then in a second step analyze the spatial dependence.

We may extend this approach to handle multivariate extremes.  Let $(Y_{i 1}, ..., Y_{i D})$, $i=1, ..., n$ be independent and identically distributed replicates of a $D-$dimensional random vector and let $M_{n} = (M_{n 1}, ..., M_{n D})$ be the vector of componentwise maxima, where $M_{n d} = \max(Y_{1 d}, ..., Y_{n d})$ for $d=1, ..., D$.  A non-degenerate limit for $M_{n}$ exists if there exist sequences $a_{nd} > 0$ and $b_{nd}$, $d=1, ..., D$ such that
\[
\lim_{n \rightarrow \infty} pr \left(\frac{M_{n1}-b_{n1}}{a_{n1}} \leq y_{1}, ..., \frac{M_{nD}-b_{nD}}{a_{nD}} \leq y_{D} \right) = G(y_{1}, ..., y_{D}).
\]
Then $G$ is a multivariate extreme value distribution, and is max-stable in the  sense that for any $n \ge 1$ there exist sequences $A_{nd}>0$, $B_{nd}$, $d=1, ..., D$ such that
\[
G^{n}(y_{1}, ..., y_{D}) = G(A_{n1}y_{1} + B_{n1}, ...,A_{nD}y_{D} + B_{nD}).
\]
The marginal distributions of a multivariate extreme value distribution are necessarily univariate Generalized Extreme Value distributions.

\subsection{Max-stable Processes}
A common method of modeling spatial extremes is through max-stable processes, which arise as an infinite dimensional generalization of multivariate extreme value theory.  Let $Z(x), x \in X \subseteq \mathbb{R}^{p}$ be a stochastic process.  If for all $n \ge 1$, there exist sequences $a_{n}(x), b_{n}(x)$ for some $x_{1}, ..., x_{D} \in X$ such that
\[
pr^{n} \left(\frac{Z(x_{d}) - b_{n}(x_{d})}{a_{n}(x_{d})} \leq z(x_{d}), d = 1, ..., D \right) \rightarrow G_{x_{1}, ..., x_{D}}(z(x_{1}), ..., z(x_{D})),
\]
then $G_{x_{1}, ..., x_{D}}$ is a multivariate extreme value distribution.  If the above holds for all possible $x_{1}, ..., x_{D} \in X$ for any $D \ge 1$, then the process is a max-stable process.  Without loss of generality one may assume that a max-stable process has unit-Fr\'{e}chet margins.  In practice, this is achieved by estimating the three Generalized Extreme Value parameters at each location $x_{d}$, and then transforming the data.

Max-stable processes are often shown in spectral representation, and are often constructed as follows:  Let $Y(x)$ be a non-negative stationary process on $\mathbb{R}^{p}$ such that $E(Y(x)) = 1$ at each $x$.  Let $\Pi = \{ s_{i} \} _{i \in \mathbb{N}}$ be a Poisson process on $\mathbb{R}_{+}$ with intensity $d s/s^{2}$.  If $Y_{i}(x)$ are independent replicates of $Y(x)$, then
\[
Z(x) = \max_{i} s_{i} Y_{i}(x), \hspace{5mm} x \in X
\]
is a stationary max-stable process with unit Fr\'{e}chet margins \citep{dehaan84}.  From this, the joint distribution may be represented as
\[
pr(Z(x) \leq z(x), x \in X) = \exp \left\{-E \left(\sup_{x \in X}  \frac{Y(x)}{z(x)} \right)\right\}.
\]

\begin{figure}
\begin{center}
\includegraphics[width=3in, angle=-90]{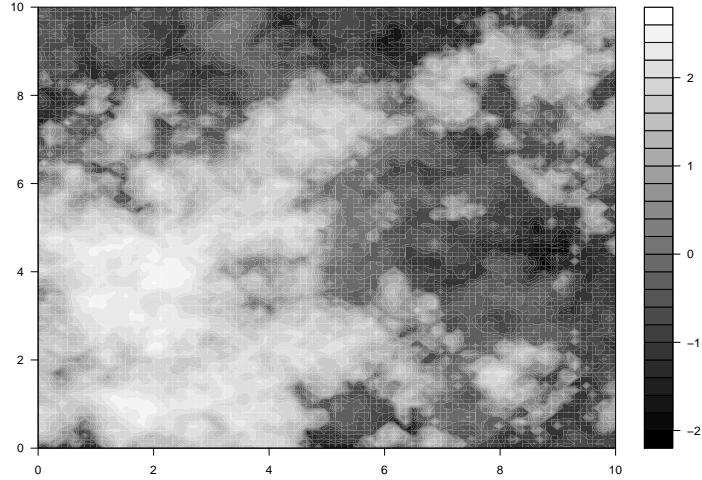}
\hspace{3mm}
\caption{Extremal Gaussian process with Whittle-Mat\'{e}rn correlation and parameter $\phi = (c_{1}, c_{2}, \nu) = (1, 3, 1)$}
\label{fig:schlather2}
\end{center}
\end{figure}

\citet{schlather02} introduced a flexible set of models for max-stable processes, termed extremal Gaussian processes.  He considered a  stationary Gaussian process $Y(x)$ on $\mathbb{R}^{p}$ with correlation function $\rho(\cdot)$ and finite mean $\mu = E \max(0, Y(x)) \in (0, \infty)$.  Let $s_{i}$ be a Poisson process on $(0, \infty)$ with intensity measure $\mu^{-1}s^{-2}ds$.  Then
\[
Z(x) = \max_{i} s_{i} \max(0, Y_{i}(x))
\]
is a stationary max-stable process with unit-Fr\'{e}chet margins.  The bivariate distribution function is
\begin{equation}
pr(Z_{1} \leq z_{1}, Z_{2} \leq z_{2}) = \exp \left(- \frac{1}{2} \left[\frac{1}{z_{1}} + \frac{1}{z_{2}} \right] \left[1 + \left\{1 - 2(\rho(h)+1)\frac{z_{1}z_{2}}{(z_{1}+z_{2})^{2}}\right\}^{1/2} \right] \right)
\label{eq:sch}
\end{equation}
where $\rho(h)$ is the correlation of the underlying Gaussian process $Y(x)$ and $h=||x_{1}-x_{2}||$.  The correlation is chosen from one of the valid families of correlations for Gaussian processes.  A few common choices are Whittle-Mat\'{e}rn, \[
\rho(h) = c_{1} \frac{2^{1-\nu}}{\Gamma(\nu)} \left(\frac{h}{c_{2}}\right)^{\nu}K_{\nu}\left(\frac{h}{c_{2}}\right), \hspace{3mm} 0 \leq c_{1} \leq 1, c_{2} > 0, \nu >0,
\]
Cauchy,
\[
\rho(h) = c_{1} \left\{1+\left(\frac{h}{c_{2}}\right)^{2} \right\}^{-\nu},  \hspace{3mm} 0 \leq c_{1} \leq 1, c_{2} > 0, \nu >0,
\]
and powered exponential
\[
\rho(h) = c_{1} \exp \left\{-\left(\frac{h}{c_{2}}\right)^{\nu}\right\} \hspace{3mm} 0 \leq c_{1} \leq 1, c_{2} > 0, 0 < \nu \leq 2,
\]
where $c_{1}, c_{2}$ and $\nu$ are the nugget, range, and smooth parameters, $\Gamma$ is the gamma function and $K_{\nu}$ is the modified Bessel function of the third kind with order $\nu$.  It is common to fix the nugget as $c_{1}=1$, which forces $\rho(h) \rightarrow c_{1} = 1$ as $h \rightarrow 0$.  This is a reasonable assumption for many environmental processes.  Throughout the remainder of this paper, we will use the Bayesian notation $\phi$ to refer to the parameters $(c_{2}, \nu)$ of the correlation function, and write $\rho(h) = \rho(h; \phi)$ to serve as a reminder that the target of our method is the function $\rho(\cdot)$.  Figure \ref{fig:schlather2} shows one realization of a process with the Whittle-Mat\'{e}rn correlation function.

There are other classes of max-stable processes which the method developed in this paper could be applied to.  One is the Smith class (Smith 1990, unpublished manuscript), formed as follows: Let ${(s_{i}, x_{i}), i \geq 1}$ denote the points of a Poisson process on $(0, \infty) \times \mathbb{R}^{p}$ with intensity measure $s^{-2}ds dx$.  Take $f(x)$ to be the multivariate Gaussian density function,
\[
f(x) = (2\pi)^{-p/2}|\Sigma|^{-1/2} \exp \left(- \frac{1}{2}x^{T}\Sigma^{-1}x \right).
\]
Then $Z(x) = \mbox{max}_{i} s_{i}f(x-x_{i})$ is a max-stable process with unit-Fr\'{e}chet margins.  Other processes are taken as $Z_{i}(x) = \exp \left[\epsilon_{i}(x) - 1/2\sigma^{2}\right]$ where $\epsilon_{i}(x)$ is a stationary Gaussian process with mean 0 and variance $\sigma^{2}$, or $Z_{i}(x) = \exp \left[\epsilon_{i}(x) - 1/2\sigma^{2}(x)\right]$ where $\epsilon_{i}(x)$ is a Gaussian process with stationary increments, mean 0, and $\mbox{var}(\epsilon(x)) = \sigma^{2}(x)$ \citep{kabluchko09}.  This process is also called a Brown-Resnick process \citep{brown77}.  We elected to concentrate on the Schlather family because it has been widely studied in recent years \citep{blanchet11} and its simulations appear realistic for practical datasets.

Two key features of max-stable processes are repeated.  First, it is possible to simulate max-stable processes through a point process approach \citep{schlather02}.  Second, only the marginal and bivariate distribution functions can be written in closed-form, so the joint likelihood function is unavailable.  These two statements taken together motivate the use of approximate Bayesian computing for modeling dependence in spatial extremes.

\subsection{The Extremal Coefficient}
Let $Z(x)$ be a stationary, isotropic max-stable random field with unit-Fr\'{e}chet margins.  For $D$ fixed locations, the joint distribution function can be written as
\[
pr(Z(x_{1}) \leq z_{1}, ..., Z(x_{D}) \leq z_{D}) = \exp\left\{-V(z_{1}, ..., z_{D})\right\}
\]
where $V(z_{1}, ..., z_{D})$ is the exponent measure first described by \citet{pickands81}.  Max-stability implies that for all $N$,
\[
pr(Z_{1} \leq z_{1}, ..., Z_{D} \leq z_{D})^{N} = \exp\{-N V(z_{1}, ..., z_{D})\} = \exp\{-V(z_{1}/N, ..., z_{D}/N)\}.
\]
This is the homogeneity property of the exponent measure.  For $D \leq 2$ locations this function $V(\cdot)$ can be written out explicitly, but for $D \ge3$ it cannot.  If we further consider the joint distribution of $D$ locations evaluated at the same value $z$, we get
\[
pr(Z(x_{1}) \leq z, ..., Z(x_{D}) \leq z) = \exp\left\{-\frac{\theta(x_{1}, ..., x_{D})}{z}\right\}
\]
where $\theta(x_{1}, ..., x_{D}) = V(1, ..., 1)$ is the extremal coefficient for the $D$ locations.  The coefficient takes values between 1 and $D$, with a value of 1 corresponding to complete dependence among the locations, and a value of $D$ corresponding to complete independence.  The value can be thought of as the number of effectively independent locations among the $D$ under consideration.

Many results are available for the pairwise extremal coefficient, which arises when considering any pair of locations,
\[
Pr(Z(x_{1}) \leq z, Z(x_{2}) \leq z) = \exp\left(-\frac{\theta(x_{1}, x_{2})}{z}\right)
\]
Since the bivariate distribution functions are available in closed form equation (\ref{eq:sch}) for the Schlather process, one may write out the pairwise extremal coefficients explicitly as
\begin{eqnarray}
\theta(h) &=& 1 + \left\{\frac{1 - \rho(h; \phi)}{2}\right\}^{1/2}
\label{eq:alpha1}
\end{eqnarray}
where $h=||x_{1} - x_{2}||$.  One may estimate the pairwise extremal coefficients directly from the data, and then through those estimates obtain an estimate of $\rho(\cdot)$.  Smith (Smith 1990, unpublished manuscript) and \citet{coles99} proposed an estimate of the pairwise extremal coefficients as follows.  First, assume that the field $Z(\cdot)$ has been transformed to unit-Fr\'{e}chet.  This means that $1/Z(\cdot)$ is unit exponential, and $1/\mbox{max}(Z(x_{1}), Z(x_{2}))$ is exponential with mean $1/\theta(x_{1},x_{2})$.  A simple estimator then is
\begin{equation}
\hat{\theta}(x_{1}, x_{2}) = \frac{n}{ \sum_{i=1}^n 1/\mbox{max}(z_{i}(x_{1}), z_i(x_{2}))}
\label{eq:alpha2hat}
\end{equation}
where $i$ is the index for the block.  In this paper, we move beyond the pairwise extremal coefficient and also focus on the tripletwise extremal coefficient, which is defined for any triplet of locations in the relation
\begin{equation}
pr(Z(x_{j}) \leq z, Z(x_{k}) \leq z, Z(x_{l}) \leq z) = \exp\left\{-\frac{\theta(x_{j}, x_{k}, x_{l})}{z}\right\}.
\label{eq:alpha}
\end{equation}
Since the trivariate distribution function for $Z(\cdot)$ is unavailable, so too is any closed-form expression for $\theta(x_{j}, x_{k}, x_{l})$.  However, following the same argument as in the pairwise case, we may estimate the coefficients using the estimator
\begin{equation}
\hat{\theta}(x_{j}, x_{k}, x_{l}) = \frac{n}{ \sum_{i=1}^{n} 1/\mbox{max}(z_{i}(x_{j}), z_{i}(x_{k}), z_{i}(x_{l}))}
\label{eq:phihat}
\end{equation}
where $i$ is the index for the block.  These estimated triplets will serve a key function in the approximate Bayesian computing algorithm.  This argument may be extended to estimate all $k$-point extremal coefficients for any collection of $k$ locations.

\section{Approximate Bayesian Computing}

Approximate Bayesian computing (ABC), also called likelihood-free computing, aims to approximate a posterior distribution when the true likelihood is either unavailable or computationally prohibitive.  Only simulations from the (unavailable) likelihood are needed.  Approximate Bayesian computing originated in population genetics \citep{fu97, tavare97, pritchard99, beaumont02}, and has since branched out into many areas, including extremes \citep{bortot07}.  Our target is the posterior distribution $\pi(\phi \mid z) \propto f(z \mid \phi)\pi(\phi)$, where $z$ is the observed data.  \citet{sisson10} described how the ABC method facilitate the computation by introducing an auxiliary parameter $z'$ (a simulated dataset) on the same space as observed data $z$.  Thus we are actually computing
\[
\pi_{ABC}(\phi, z' \mid z) \propto \pi(z \mid z', \phi) \pi(z' \mid \phi) \pi(\phi).
\]
Integrating out the simulated dataset yields the target posterior of interest
\[
\pi_{ABC}(\phi \mid z) \propto \pi(\phi) \int_{Z} \pi(z \mid z', \theta)\pi(z' \mid \phi)\,dz'.
\]
When $\pi(z \mid z', \phi)$ is exactly a point mass at the point $z'=z$ and zero everywhere else, the posterior is recovered exactly.  This is likely to occur with probability 0 (for continuous data), or probability close to zero (for discrete but high dimensional data), so in practice the form is usually taken to be
\[
\pi(z \mid z', \phi) =  \frac{1}{\epsilon} K \left(\frac{|S(z') - S(z)|}{\epsilon} \right)
\]
where $K$ is a kernel density function, and $S$ is a summary statistic.  The intractable likelihood is weighted in regions where $S(z') \approx S(z)$.  When $S$ is a sufficient statistic and in the limit as $\epsilon \rightarrow 0$, we have $\lim_{\epsilon \rightarrow 0} \pi_{ABC}(\phi \mid z) = \pi(\phi \mid z)$.  The most familiar form of this is
\[
\pi(z \mid z', \theta) \propto 1 \hspace{2mm} \mbox{if} \hspace{2mm} d(S(z'), S(z)) \leq \epsilon
\]
which occurs when $K$ is a uniform density kernel on an interval.  With these concessions, the most basic implementation is as follows:
\begin{enumerate}
\item Draw a candidate parameter $\phi' \sim \pi(\phi)$ from a prior
\item Simulate data from $ f(Z' \mid \phi')$
\item Accept $\phi'$ if $d(s, s') \leq \epsilon$ and return to step 1.
\end{enumerate}

This is the form shown by \citet{marjoram03}, using the uniform kernel (in this paper we focus on the uniform kernel exclusively, and drop the kernel notation $K(\cdot)$ in favor of the more straightforward acceptance notation $d(S(z'), S(z)) \leq \epsilon$).  Choosing the quantities $S$, $d$, and $\epsilon$ is necessarily a trade-off between accuracy of the approximation and computational efficiency.  In many applications, the challenge is that the chosen summary $S$ must be highly informative of the parameter, while at the same time $pr(d(S(z), S(z')) \leq \epsilon)$ must be large enough for the entire algorithm to be computable.  The algorithm produces an independent and identically distributed sample drawn from $\pi(\phi \mid d(S(z), S(z')) \leq \epsilon)$, and the following two limits show the role of the threshold $\epsilon$:
\begin{enumerate}
\item If $\epsilon \rightarrow \infty$, then $\pi(\phi \mid d(S(z), S(z')) \leq \epsilon) \rightarrow \pi(\phi)$
\item If $\epsilon \rightarrow 0$, then $\pi(\phi \mid d(S(z), S(z')) \leq \epsilon) \rightarrow \pi(\phi \mid S(z))$
\end{enumerate}
For large values of $\epsilon$, nearly all draws from the prior will lead to acceptances, thus the approximated posterior mirrors the prior.  As the threshold is reduced, the approximation more closely resembles $\pi(\phi \mid S(z))$.  When $S$ is a sufficient statistic, this is the exact posterior.  However, $S$ is unlikely to be sufficient in practice, and is often chosen to be highly informative of the parameter $\phi$.  The remainder of this section will discuss three particular implementations of approximate Bayesian computing for spatial extremes.  In each case, the particular motivation and definition for the summary will be discussed.

\subsection{The Madogram Method}
Let $Z(x)$ be a stationary, isotropic max-stable random field with Generalized Extreme Value margins with $\xi < 1$.  The madogram is defined as:
\[
m(h) = \frac{1}{2} E | Z(x+h) - Z(x) |,
\]
and its natural estimator is defined as
\begin{equation}
\hat{m}(h) = \frac{1}{2n} \sum_{i=1}^{n} |z_{i}(x) - z_{i}(x+h)|,
\label{eq:mhat1}
\end{equation}
where $z_{i}(x)$ is the realization of the $i^{th}$ observed process at position $x$.  This estimator is unbiased.  \citet{cooley06} showed the relationship between the madogram and the extremal coefficient $\theta(h)$.  If the Generalized Extreme Value shape parameter $\xi < 1$, then the madogram $m(h)$ and extremal coefficient $\theta(h)$ verify
 \[
   \theta(h) = \left\{
     \begin{array}{lr}
       u_{\beta} + \frac{m(h)}{\Gamma(1-\xi)} & \mbox{if }\xi<1 \mbox{ and } \xi \neq 0\\
       \vspace{3mm}
       \exp \left(\frac{m(h)}{\sigma}\right) & \mbox{ if } \xi=0,
     \end{array}
   \right.
\]
where $u_{\beta} = \left(1 + \xi \frac{u - \mu}{\sigma}\right)_{+}^{1/\xi}$ and $\Gamma(\cdot)$ is the Gamma function.  Note in particular that for unit-Gumbel margins (with $\xi=0$ and $\sigma=1$), we have the simple relationship $m(h) = \log{\theta(h)}$.  We will exploit this simple relationship by first transforming all margins of a max-stable process to unit-Gumbel (and not the usual unit-Fr\'{e}chet).  This is easily done by taking the log of data with unit-Fr\'{e}chet margins.

Thus assuming that the marginal parameters of the process are known, the estimator of the madogram is unbiased, and we have a closed-form expression for the madogram as a function of the underlying correlation $\rho(h; \phi)$, which is the target of our method.  We can naturally define a residual as
$e(h) = \hat{m}(h) - \log{\theta(h)}$.  Thus, for the Schlather model, plugging in equations (\ref{eq:alpha1}) and (\ref{eq:mhat1}) we obtain residuals
\[
e(h) =\frac{1}{2n} \sum_{i=1}^{n} |z_{i}(x) - z_{i}(x+h)| - \log \left\{1 + \left(\frac{1 - \rho(h; \phi)}{2}\right)^{1/2} \right\}.
\]
The parameter value which minimizes the sum of squared residuals is the ordinary least squares estimator, equal to
\begin{equation}
\hat{\phi}_{OLS} = \mbox{argmin}_{\phi} \sum_{h} e(h)^{2}.
\label{eq:ols}
\end{equation}
The summary statistic $S$ is chosen to be the ordinary least squares fit to the madogram, subject to the constraint that it be a valid madogram.  Mathematically, this is
\begin{equation}
S = \log\{\theta(h; \hat{\phi}_{OLS})\}.
\label{eq:s1}
\end{equation}
\begin{figure}
\begin{center}
\includegraphics[width=3in, angle=-90]{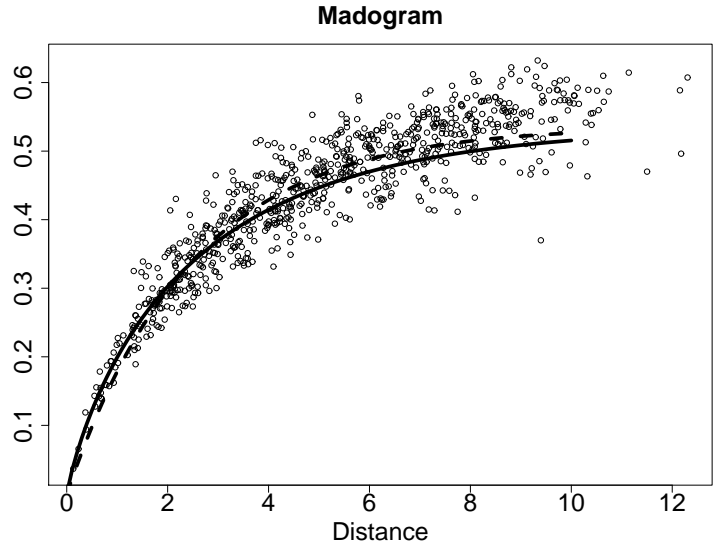}
\caption{Example of a madogram (solid line), estimate (points), and ordinary least squares fit (dotted line).  The entire dotted line is the summary statistic, defined by parameter $\hat{\phi}_{OLS}$.}
\label{fig:madogram}
\end{center}
\end{figure}
The procedure for utilizing the summary statistic is as follows.  For observed data $Z$, the madogram is estimated and the OLS fit is obtained using equation (\ref{eq:ols}).  Then the summary statistic $S=s$ is computed using equation (\ref{eq:s1}).  For each successive iteration of the approximate Bayesian computing algorithm, a simulated data set $Z'$ is obtained from parameter $\phi' \sim \pi(\phi)$.  The madogram is estimated and an OLS fit to the madogram $S'=s'$ is obtained.  What remains is some means of computing the distance between $s$ and $s'$.  We have chosen this as
\[
d(s, s') = \int |s(h) - s'(h)| \,dh.
\]
The integral of the absolute differences between the two curves $s$ and $s'$ is computed numerically, and taken as a measure of the distance between $s$ and $s'$.  The final step in approximate Bayesian computing is to accept $\phi'$ for the posterior if $d(s, s') \leq \epsilon$ for some suitably chosen $\epsilon$.

The output of this is a collection of $M$ particles $\phi'_{1}, ..., \phi'_{M}$ which are taken to be a sample from the approximated posterior.  From this, we computed the correlation function $\rho(h; \phi)$ evaluated for each $\phi'_{m}$.  The analog of a posterior mean in this setting is the pointwise mean of all accepted functions,
\begin{equation}
\hat{\rho}(h) = \frac{1}{M} \sum_{m=1}^{M} \rho(h; \phi'_{m}).
\label{eq:phiM}
\end{equation}
We use the pointwise mean when evaluating the performance in a simulation study.

\subsection{The Pairwise Extremal Coefficient Method}
This approach is very similar to the preceding madogram approach, but instead of fitting a smooth curve to the madogram we fit the curve directly to the pairwise extremal coefficients.  We define the residual as $e(h) = \hat{\theta}(h) - \theta(h)$.  Plugging in equations (\ref{eq:alpha1}) and ({\ref{eq:alpha2hat}), the parameter value which minimizes the sum of residuals is equal to
\[
\hat{\phi}_{OLS} = \mbox{argmin}_{\phi} \sum_{h} e(h)^{2}.
\]
The summary statistic $S$ is chosen to be the ordinary least squares fit to the madogram, subject to the constraint that it be a valid madogram.  Mathematically, this is
\begin{equation}
S = \theta(h; \hat{\phi}_{OLS}).
\label{eq:s2}
\end{equation}
The remainder proceeds exactly as in the madogram method, using the summary shown in equation (\ref{eq:s2}).

\subsection{The Tripletwise Extremal Coefficient Method}
Both the madogram approach and the pairwise extremal coefficient approach rely on pairs of locations.  This is also true for the composite likelihood approach \citep{padoan10}.  A natural improvement is an approximate Bayesian computing method which moves beyond pairs and considers higher order k-tuples.  The use of triplets was explored by \citet{genton11}, but only for the Smith model \citep{smith90}, a small subset of max-stable processes that does not include the Schlather model.  In this section we use the estimated triplet extremal coefficients from equation (\ref{eq:phihat}) as the basis for the summary statistic $S(\cdot)$, and thus utilize information from triplets in the estimation of Schlather max-stable processes.

The number of unique sets of triplets in a set of data with $D$ locations is ${D \choose 3} = \frac{D (D-1) (D-2)}{6}$, which grows quite rapidly as $D$ increases.  For example, with only $D=20$ locations we have 1140 unique triplets.  This combinatorial explosion as $D$ increases poses a problem for an approximate Bayesian computing approach.  Higher dimensional summaries can only decrease the probability of acceptances, which may quickly leave an approach uncomputable in any practical sense.  On the other hand, the uncertainty in estimating a single triplet extremal coefficient using equation (\ref{eq:phihat}) can be quite large (as compared with the known bounds [1,$D$]), so there is a natural desire to group estimates into homogeneous groups and take averages to reduce the uncertainty in estimation.  The idea then is to group the ${D \choose 3}$ triplets into $K$ groups, which are ideally homogeneous within groups, heterogeneous across groups, and all such that $K \ll {D \choose 3}$.

To reduce the dimension of the summary, we group these ${D \choose 3}$ triplets into $K$ groups using Ward's method \citep{ward63}.  This method only requires a measure of distance between items, and the number of groupings.  A triplet of locations is a triangle between 3 points, which produces 3 Euclidean distances $A = (a_{1}, a_{2}, a_{3})$.  To measure the distance between two triplets $A$ and $B$, we take
\begin{equation}
dist(A, B) = \min_{\pi} \sum_{i} |a_{i} - b_{\pi(i)}|
\label{eq:dist}
\end{equation}
where $\pi(i)$ is a permutation of $\{1, 2, 3\}$.  Two sets of triplets $A$ and $B$ with identical lengths but different rotations, translation, and reflection would give a distance measure of zero.  Such two sets should also have the same theoretical tripletwise extremal coefficients since the underlying field is isotropic and stationary.  On the other hand, as two triangles become more dissimilar in their respective lengths, the distance measure will increase.  Thus the clustering is based entirely on the geometry of the locations, and not on the actual estimates of the tripletwise extremal coefficients.

For a data set with $D$ locations, the first step is to compute an upper triangular dissimilarity matrix of size ${D \choose 3}$ by ${D \choose 3}$ which contains all distances computed using equation (\ref{eq:dist}).  In our simulations based on $D=20$ locations, we chose to group the triplets into $K=100$ clusters.  This was selected to achieve a balance between maintaining within-group homogeneity and ensuring that enough triplets fall into each group to reduce variability in group averages.  Ward's method is a hierarchical algorithm which first assigns each item to its own cluster, and then merges two clusters chosen to minimize the overall increase in the sum of squares (which is the sum of squared distances from each item to its cluster center).  Thus, the sum of squares begins at zero, and Ward's method proceeds by merging items which would result in the smallest increase.  In our setting this clustering only needs to be done once, since all of the simulated draws will be at the same locations as the observed data.  The requirement to enumerate all ${D \choose 3}$ triplets is a practical limitation to how large $D$ may be.  For values of $D$ where the dissimilarity matrix is computable, it may be time consuming to run the clustering algorithm.

The ${D \choose 3}$ triplet extremal coefficients are estimated for the observed data using equation (\ref{eq:phihat}), and then values are averaged within the $K$ clusters.  The result is the summary of the observed data, $s = (\bar{\theta}_{1}, ..., \bar{\theta}_{K})$.  Next, we begin the approximate Bayesian computing procedure. Independent draws from the prior $\phi' \sim \pi(\phi)$ are taken.  The parameter space is $\Phi = (0, \infty) \times (0, \infty)$, except when a powered exponential is used in which case it is $\Phi = (0, \infty) \times (0, 2]$.  For each draw from the prior, a max-stable process with unit-Fr\'{e}chet margins is simulated on the same locations and for the same number of years as the observed data.  We estimate all triplet extremal coefficients for this simulated data, compute $s' = (\bar{\theta}'_{1}, ..., \bar{\theta}'_{K})$, and use the sum of the absolute deviations as the distance metric $d$:
\begin{equation}
d(s, s') = \sum_{k=1}^{K} |s_{k} - s'_{k}|.
\label{eq:d}
\end{equation}
This entire process is repeated $I$ times.  The result is a collection of candidate parameter values $(\phi'_{i}, d_{i}), i=1, ..., I$, which are then filtered as $(\phi'_{i}: d_{i} \leq \epsilon)$.  This final filtration is an independent and identically distributed collection of $M$ particles drawn from $\pi(\phi \mid d(s, s') \leq \epsilon)$, which for very small $\epsilon$ may be taken as an approximation to the true posterior.  For each particle one can compute the spatial correlation function $\rho(h; \phi'_{m})$.  Standard errors are obtained by simply regarding $\rho(\phi'_{m})$ as an independent and identically distributed collection of draws from the posterior, and empirical 95\% credible intervals can be constructed.

\section{Simulation Study}
\subsection{ABC Rejection Simulations}
We study the performance of the approximate Bayesian computing algorithm for estimating the spatial dependence of a Schlather process with Whittle-Mat\'{e}rn correlation $\rho(c_{1}=1, c_{2}, \nu)$.  Simulations were conducted in \texttt{R}.  We specified uniform, independent priors on [0,10] for the range $c_{2}$ and smooth $\nu$ parameters.  This nicely spans the range of possible dependence functions on the space $X$ (see Figure \ref{fig:cor}), and is consistent with the preference for minimally informative priors.  While this prior may not be the most efficient choice, it does suffice to show the advantages of approximate Bayesian computing over the composite likelihood approach.  We make the comparison using mean square error as our measure of performance.  The simulations were all carried out for $n=100$ years of data at $D=20$ locations whose locations were drawn from a uniform distribution on a 10 by 10 grid.

For each dataset we estimated the spatial dependence using both the composite likelihood approach and the approximate Bayesian computing approach shown in equation (\ref{eq:phiM}).  Figure \ref{fig:cor} shows an example.  Approximate Bayesian computing was done with $I=1,000,000$ draws.  Due to the substantial computing time needed, we ran the simulations in parallel on 50 nodes on a research computing cluster, with each node only responsible for simulating 20,000 datasets.  In parallel, total computing time for one dataset in one model was around 8 hours for the madogram method (which contains a numeric optimization step for each iteration), but often faster for the ABC pairwise and ABC tripletwise approaches.  Given this constraint, we chose to limit the number of repetitions to only 5 replications for each model.  In all there were 6 models, therefore 30 simulation runs (in the next subsection we discuss a faster implementation with more simulations).

The output from each simulation was filtered as $(\phi'_{i}: d_{i} \leq \epsilon_{P})$, where $\epsilon_{P}$ is the $0.02\%$ percentile of $d_{i}$.  This ensures exactly 200 particles are accepted for the approximate posterior distribution for each simulation.  We found that the spacings of the ($D=20$) locations can shift the overall distribution of $d_{i}$, so for identical model specifications one may need different thresholds of $\epsilon$ to ensure enough particles are accepted.  Thus, it is better not to specify a fixed threshold $\epsilon$ but instead set as a very low percentile.

We judged relative performance of the methods based on estimating the true correlation $\rho(h; \phi_{TRUE})$, not in estimating the true parameter $\phi_{TRUE}$.  Two very different parameters $\phi_{1}$ and $\phi_{2}$ can produce similar correlations $\rho(h; \phi_{1}) \approx \rho(h; \phi_{2})$ (by moving the range and smooth parameters in opposite directions, for example).  A diffuse posterior distribution of $\phi$ can actually produce a tight posterior distribution of $\rho(h; \phi)$; we have observed that the ABC method shows this behavior.  Thus, the comparison is made between the true correlation function $\rho(h; \phi_{TRUE})$ and estimated correlation function under the various approaches: $\hat{\rho}(h) = \rho(h; \hat{\phi}_{MCLE})$ for the composite likelihood method, and for the ABC approaches the pointwise posterior mean in equation (\ref{eq:phiM}).

Mean square error was computed as a numeric approximation to
\begin{equation}
\textsc{MSE} = \int_{\{h > 0: \hspace{1mm} \rho(h; \phi_{TRUE}) \ge 0 . 1\}}  \left(\rho(h; \phi_{TRUE}) - \hat{\rho}(h)\right)^{2} \,dh
\label{eq:mse}
\end{equation}
Taking the interval over the range $\{h > 0: \hspace{1mm} \rho(h; \phi_{TRUE}) \ge 0.1\}$ focuses the comparison on the regions of higher spatial correlation, which are of greater interest.  If we computed MSE over the entire range of $\rho(\cdot)$, results of this paper would not differ in any meaningful way.  We stress that the pointwise mean in equation (\ref{eq:phiM}) is used only to compare the ABC methods with the composite likelihood approach in the simulation study.  When the ABC approach is used alone in practice, one would use the full approximate posterior to handle prediction, credible intervals, and assess uncertainty.  The application at the end of this manuscript shows an example.

\begin{figure}
\begin{center}
\includegraphics[width=5in]{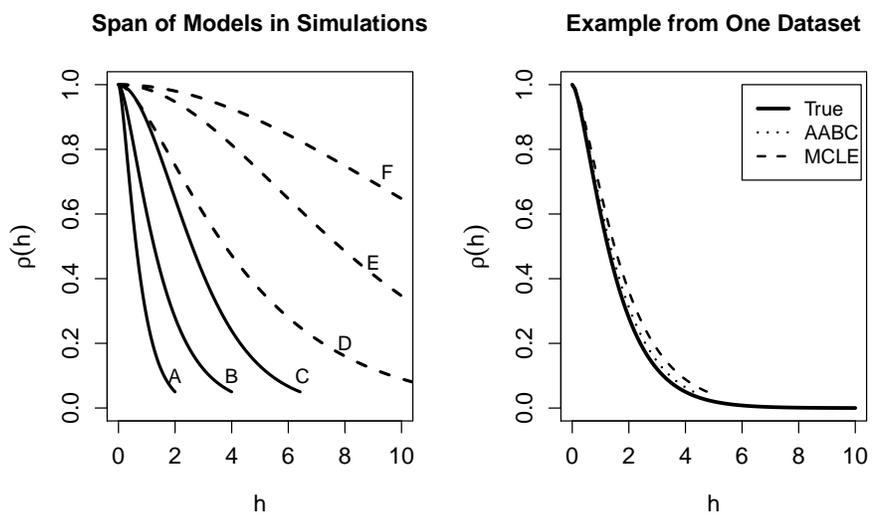}
\caption{Left: Span of models included in simulation study.  Models range from short-range dependence (A) to long-range dependence (F) on the scale of simulated data.  In the three models labeled A, B, and C (solid lines) the adaptive ABC tripletwise approach outperformed MCLE.   Right: Example of approximate Bayesian computing (dashed line) and maximum composite likelihood (dotted line) estimates from one model run, as compared to the true model (solid line).}
\label{fig:cor}
\end{center}
\end{figure}
\begin{table}
\begin{center}
\caption{Mean Square Error using equation (\ref{eq:mse}) for both the composite likelihood and three approximate Bayesian computing methods.  Reported values are averages from 5 simulations of each of the six models.  Standard error estimates are shown in brackets.  The values reported have order of magnitude is $10^{-4}$.  The final column shows the relative reduction in $\textsc{MSE}$ when using the approximate Bayesian computing tripletwise method, as compared to the composite likelihood method.}
\vspace{3mm}
\small
\begin{tabular}{c c c c c | c}
\hline
 & ABC & ABC & ABC & Composite & Reduction: ABC Tripletwise \\
Model & Madogram & Pairwise & Tripletwise & Likelihood & vs. Composite Likelihood\\
\hline

A $(c_{2}=0 . 5, \nu=1)$ & 99 [68]  & 45 [27] & 15 [9]  & 26 [13] &$ 40 . 1\%$\\
B $(c_{2}=1, \nu=1)$ & 207 [81]  & 262 [102]  & 125 [77]  & 146 [121]  & $14 . 5\%$\\
C $(c_{2}=1, \nu=3)$ & 314 [104]  & 272 [37]  &  103 [46] & 140 [43] & $26 . 6\%$\\
D $(c_{2}=3, \nu=1)$ & 98 [38] & 284 [155] & 99 [69]  & 163 [74] &$ 39.4\%$\\
E $(c_{2}=3, \nu=3)$ & 385 [208] & 555 [241] & 287 [100] & 283 [105] & $-1 . 3\%$ \\
F $(c_{2}=5, \nu=3)$ & 269 [92] & 645 [174] & 70 [41] & 485 [374] &$ 85 .6 \%$\\
\end{tabular}
\end{center}
\end{table}

\normalsize

Results are shown in Table 1.  The first two columns show the performance of the two pairwise ABC methods, the madogram and extremal pairwise coefficient methods, followed by the ABC tripletwise and MCLE approaches.  The average MSE over 5 runs is lowest for the ABC tripletwise in 5 of 6 models, and essentially tied with composite likelihood in the sixth.  Within each model specification, having only five repetitions means standard error estimates remain large, and it is difficult to make firm conclusions.  However, when viewed as a whole, these 30 runs do show evidence in favor of the ABC tripletwise approach.  The ABC tripletwise method outperformed the ABC pairwise method in 25 out of 30 of the model runs, and it outperformed the composite likelihood approach in 19 out of 30 runs.  The ABC tripletwise method also gave the best estimate out of the four methods in 16 out of 30 runs, whereas the composite likelihood was best in only 8 of 30 runs.  We found the greatest correlation in performance between the ABC madogram and ABC pairwise methods (0.613), as expected, since these two methods are the most similar and utilize essentially the same information in the summary statistics.  These findings motivated the larger simulation study discussed in the next section.

\subsection{Adaptive ABC Simulations}
Motivated by the promising results for the ABC tripletwise approach, we carried out a second simulation study using a more computationally efficient adaptive approximate Bayesian computing (AABC) algorithm to more closely compare the performance of the ABC tripletwise and MCLE approaches.  The aim of this simulation study was to increase computational efficiency and the number of simulations per model, but avoid the use of parallel computing.  \citet{beaumont09} describe the adaptive algorithm in detail.  It is a sequential algorithm which uses ABC rejection sampling to produce a first approximation, and then re-samples from this first approximation for second round of ABC rejection sampling to produce a subsequent approximation.  This allows the second stage of ABC to sample more efficiently, thus increasing the efficiency of the algorithm.  We implemented the AABC algorithm exactly as described by \citet{beaumont09}.  Specifically, the steps were to:
\begin{enumerate}
\item Run the ABC rejection algorithm exactly as described in the section 4.1 but with $I=100,000$ simulations to produce a first approximation $\phi^{(1)}_{1}, ..., \phi^{(1)}_{J}$ (the $J=500$ particles filtered as $(\phi'_{i}: d_{i} \leq \epsilon_{P})$, where $\epsilon_{P}$ is the $0.5\%$ percentile of $d_{i}$)
\item Compute $\Omega$ as twice the empirical variance of $\phi^{(1)}_{1}, ..., \phi^{(1)}_{J}$.
\begin{enumerate}
\item Resample a particle $\phi^{*}$ from $\phi^{(1)}_{1}, ..., \phi^{(1)}_{J}$
\item Mutate using kernel $K(\phi' \mid \phi^{*}) = \mathcal{N}(\phi^{*}, \Omega)$
\item Simulate $Z' \mid \phi'$, compute summary $s'$ and distance $d(s, s')$ as before
\item (Repeat 100,000 times)
\end{enumerate}
\item Filter the 100,000 particles as $(\phi'_{i}: d_{i} \leq \epsilon_{P})$, where $\epsilon_{P}$ is the 0.5\% percentile, ensuring exactly 500 particles are accepted.  Call these $\phi^{(2)}_{m}, m=1, ..., 500$.
\item For accepted particle $\phi^{(2)}_{m}$ compute weight
\[w_{m} \propto \frac{1}{\sum_{j=1}^{J} \frac{1}{J} \cdot \mathcal{N} (\phi^{(2)}_{m} \mid \phi^{(1)}_{j},\Omega)}
\]
 where $\mathcal{N}(\cdot \mid \phi^{(1)}_{j}, \Omega)$ is the density of a multivariate normal with mean $\phi^{(1)}_{j}$ and variance $\Omega$ evaluated at the point $\phi^{(2)}_{m}$.
\end{enumerate}

The only additional consequence of this adaptive algorithm is that we have produced a weighted sample from the approximate posterior, and thus have to modify our estimate of the correlation function from equation (\ref{eq:phiM}) to now be

\begin{equation}
\hat{\rho}(h) = \frac{1}{M} \sum_{m=1}^{M} w_{m} \cdot \rho(h; \phi^{(2)}_{m}).
\label{eq:phiM2}
\end{equation}

Results for MCLE and the AABC approximation are shown in Table 2.  The more efficient AABC approach could be run without the use of any parallel computing, freeing up nodes, which allowed for 30 runs in each model (thus $6 \cdot 30 = 180$ simulations in total).  We chose an initial sampling of 100,000 and a re-sampling of 100,000 to keep the computational cost to around 8 hours per run.  This means the performance of AABC as shown in Table 2 is roughly what a user might expect when analyzing a dataset on a single computer in a single day.  The AABC method resulted in a lower MSE for the three short range processes (A, B, and C) but a larger MSE for the three longer range processes (D, E, and F).  Clearly, the adaptive ABC approach was not shown to outperform MCLE for all of the models, but there is a clear statistical benefit for the short-range processes.  We discuss this more in section 6.

\begin{table}
\begin{center}
\caption{Mean Square Error of the pointwise mean correlation function estimates (taken with respect to the true correlation function) for the MCLE and AABC methods.  Reported values are averages from 30 simulations of each of the six models.  Standard error estimates are shown in brackets.  The values reported have order of magnitude is $10^{-4}$.}
\vspace{3mm}
\small
\begin{tabular}{c c c}
\hline
Model & MCLE & AABC \\
\hline
A $(c_{2}=0 . 5, \nu=1)$ & 265 [134] & 217 [23] \\
B $(c_{2}=1, \nu=1)$ & 330 [36] & 115 [33] \\
C $(c_{2}=1, \nu=3)$ & 162 [33] & 76 [17] \\
D $(c_{2}=3, \nu=1)$ & 225 [44] & 395 [26] \\
E $(c_{2}=3, \nu=3)$ & 158 [7] & 238 [11] \\
F $(c_{2}=5, \nu=3)$ & 47 [8] & 79 [6] \\
\end{tabular}
\end{center}
\end{table}

\normalsize

\section{Application to US Temperature Data}
We illustrate the methodology of the ABC tripletwise extremal coefficient approach on US temperature data in northern Texas, with the aim of modeling the acreage of cotton at risk of an October freeze.  Data on crop losses taken from the United States Department of Agriculture Risk Management Agency shows that between 1989 and 2008, Texas cotton losses caused by freezing totaled \$108,478,787.  Of these losses, fully 67.8\% (\$73,642,461) occurred in the month of October.

The data are daily minimum temperature data taken from 30 gauged sites centered around northern Texas in the United States, freely obtained from the Global Historical Climatology Network (http://www.ncdc.noaa.gov/oa/climate/ghcn-daily/) \citep{peterson97}.  All sites are between 104 and 98 degrees west longitude and 31 to 37 degrees north latitude.  We required stations have at least 90\% of daily October values for at least 90\% of the years between 1935 and 2009.  This region is shown in Figure \ref{fig:US}.  Also shown are the 58 counties which jointly comprise the four Texas agricultural districts responsible for 82.3\% of all Texas upland cotton acreage (in 2009).  For each year and location, we took the minimum daily temperature in the month of October.  The aim of the analysis was to estimate the spatial dependence of the process through $\rho(\cdot)$, and use this information to estimate the number of acres of cotton at risk of an early freeze through simulations.
\begin{figure}
\begin{center}
\includegraphics[width=5in]{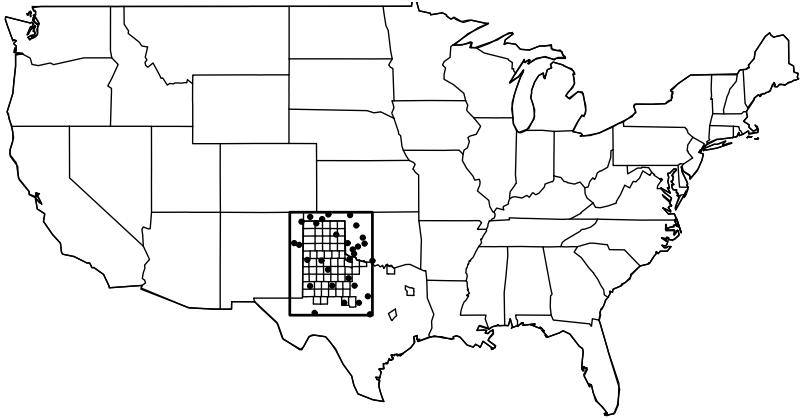}
\caption{Locations of the 30 gauged sites and 58 counties in the primary cotton growing region of Texas.}
\label{fig:US}
\end{center}
\end{figure}

\begin{figure}
\begin{center}
\includegraphics[width=5in]{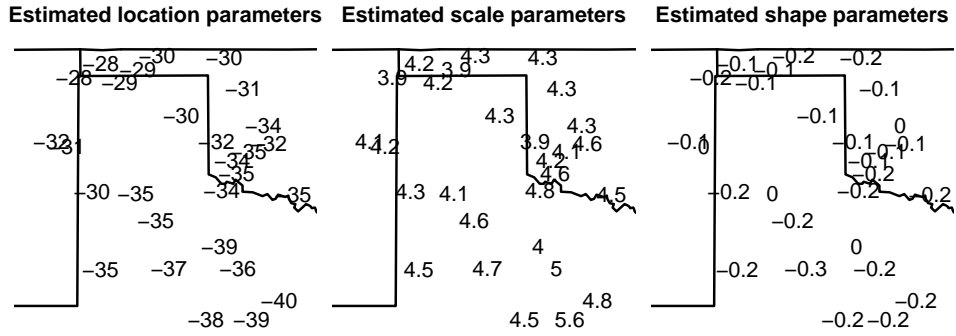}
\caption{Estimates of the Generalized Extreme Value parameters at each of the 30 gauged locations.  The location $\mu(x)$ and scale $\sigma(x)$ show spatial dependence, whereas the shape parameter $\xi(x)$ does not.  Heavy lines are US state borders.}
\label{fig:Generalized Extreme Value}
\end{center}
\end{figure}

\begin{figure}
\begin{center}
\includegraphics[width=4in]{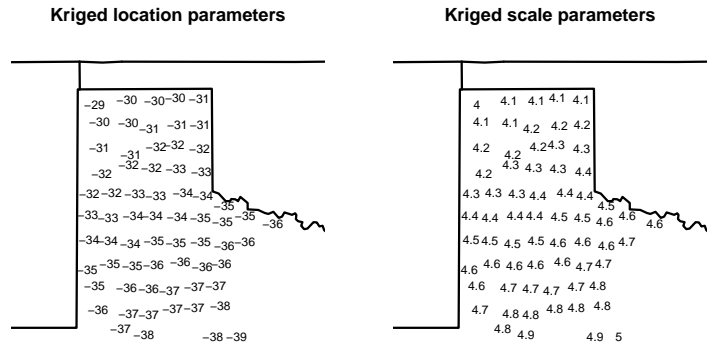}
\caption{Kriged Generalized Extreme Value parameters at each of the 58 ungauged locations.  Heavy lines are US state borders.}
\label{fig:krig}
\end{center}
\end{figure}

First we transformed data at each location to unit-Fr\'{e}chet by fitting the marginal univariate data to the Generalized Extreme Value distribution and obtained maximum likelihood estimates of the location-specific Generalized Extreme Value parameters $\mu(x), \sigma(x), \xi(x)$ for $x_{1}, ..., x_{30} \in X$.  These estimates are shown in Figure \ref{fig:Generalized Extreme Value}.  The location parameter and scale parameters were both influenced heavily by spatial location, whereas the shape parameter showed no discernible relationship to spatial placement.

Next we estimated the tripletwise extremal coefficients for the 4060 unique triplets using equation (\ref{eq:phihat}).  We clustered these into $K=100$ groups using Ward's method as described in Section 3.  Our summary statistic for the data was the average within $K$ groups, $s = (\bar{\theta}_{1}, ..., \bar{\theta}_{100})$ for $K=100$ groups.  We considered the Whittle-Mat\'{e}rn, Cauchy, and powered exponential correlation functions as possible models.  Using the standard composite likelihood information criteria for model selection \citep{padoan10}, we found the Whittle-Mat\'{e}rn to have the best fit, and selected the uniform priors for the range $c_{2}$ and smooth $\nu$ as independent uniforms U[0,10] (see section 6 for more on model selection and ABC).  For this model, this prior allows for a full range of spatial processes on the scale of the observed data in $X$.  We drew 1,000,000 draws from the prior, and ran the approximate Bayesian computing algorithm.

The threshold $\epsilon$ was set as the 0.02\% percentile of $d_{i}$.  This ensured exactly 200 particles were accepted for the approximate posterior.  The pointwise mean of these 200 accepted functions $\rho(h; \phi')$ was taken as the estimate of the correlation function.  Approximate pointwise 95\% credible intervals were estimated as the pointwise 2.5\% and 97.5\% quantiles, taken at each $h$.  This is shown in Figure (\ref{fig:result}).

The primary aim of this application was to estimate the number of acres of cotton at risk of an October freeze.  To extrapolate the max-stable process to the 58 ungauged county centroids, we obtained the county centroids from the US Census Bureau \\
(http://www.census.gov/geo/www/cenpop/county/coucntr48.html), and extrapolated location-specific Generalized Extreme Value parameters for each of these by using the standard spatial Kriging in the \texttt{R} package \texttt{fields}.  We found estimates of the location $\mu(x)$ scale $\sigma(x)$ parameters varied with location, whereas the shape parameter $\xi(x)$ showed no discernible relationship to spatial location, shown in Figure \ref{fig:Generalized Extreme Value}.  Thus, for an ungauged location $x'$ we used Kriged values of $\mu(x')$ and $\sigma(x')$, but took $\xi(x') = \frac{1}{30} \sum_{i=1}^{30} \hat{\xi}(x_{i})$, the average of the 30 estimates $\hat{\xi}(x)$ from the gauged locations (thus each $\xi(x')$ was the same for all $x'$).  Kriged values for the location and scale are shown in Figure \ref{fig:krig}.

We generated simulations from the fitted model not with the intent of matching them to observed data, but rather to calculate a distribution of cotton losses in hypothetical future years under the same climate.  We sampled $\phi$ (with replacement) from the approximate posterior and simulated a max-stable process with unit-Fr\'{e}chet margins and Whittle-Mat\'{e}rn correlation at the 58 ungauged county centroids.  We used the Kriged location-specific Generalized Extreme Value parameters to transform this back to a temperature scale at each of the 58 county centroids.  If the minimum October temperature of the county centroid fell below the chosen threshold, we assumed all acres within the county were exposed to the temperature event.  Figure (\ref{fig:count}) shows the log-density of exposed counties for 10,000 simulations at various temperature thresholds.  This method of simulation based on the approximate posterior very naturally incorporates parameter uncertainty into the quantity of interest.

Of particular interest to the insurance and agricultural communities are the number of simulations resulting in intermediate exposure, say between 0.5 and 3.5 million acres (contrasted with the all-or-nothing scenarios of 0 or the full 4.1 million acres exposed).  We find 49.8, 38.5, and 26 percent fall into this range for thresholds of 32, 30, and 28 degrees Fahrenheit. This provides evidence that these counties are not completely dependent with respect to an October freeze, and lends additional support to the idea that it may be possible to offer financial or insurance products to protect against crop losses caused by this freeze peril.  The method shown here allows for realistic estimation of a distribution of insurance losses, going beyond the empirical distribution calculated from past data.  This information could be useful to an actuary interested in calculating the expected payout associated with an insurance policy.
\begin{figure}
\begin{center}
\includegraphics[width=4in]{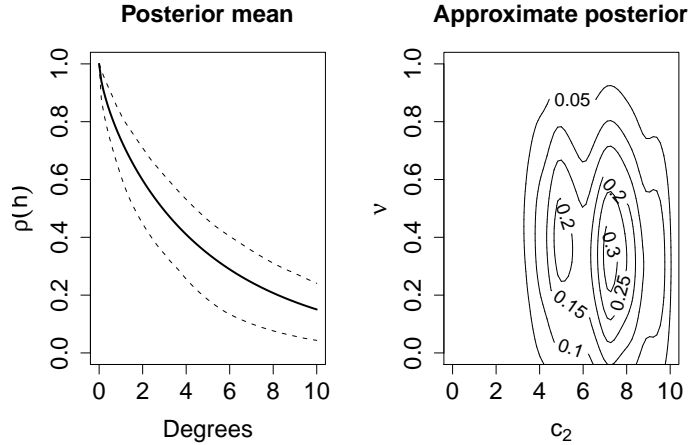}
\caption{Left: approximate Bayesian computing estimate of spatial correlation of the max-stable process, with pointwise 95\% credible interval estimates shown as dotted lines.  Right: Approximate posterior of the range and smooth parameters.}
\label{fig:result}
\end{center}
\end{figure}
\begin{figure}
\begin{center}
\includegraphics[width=3in, angle=-90]{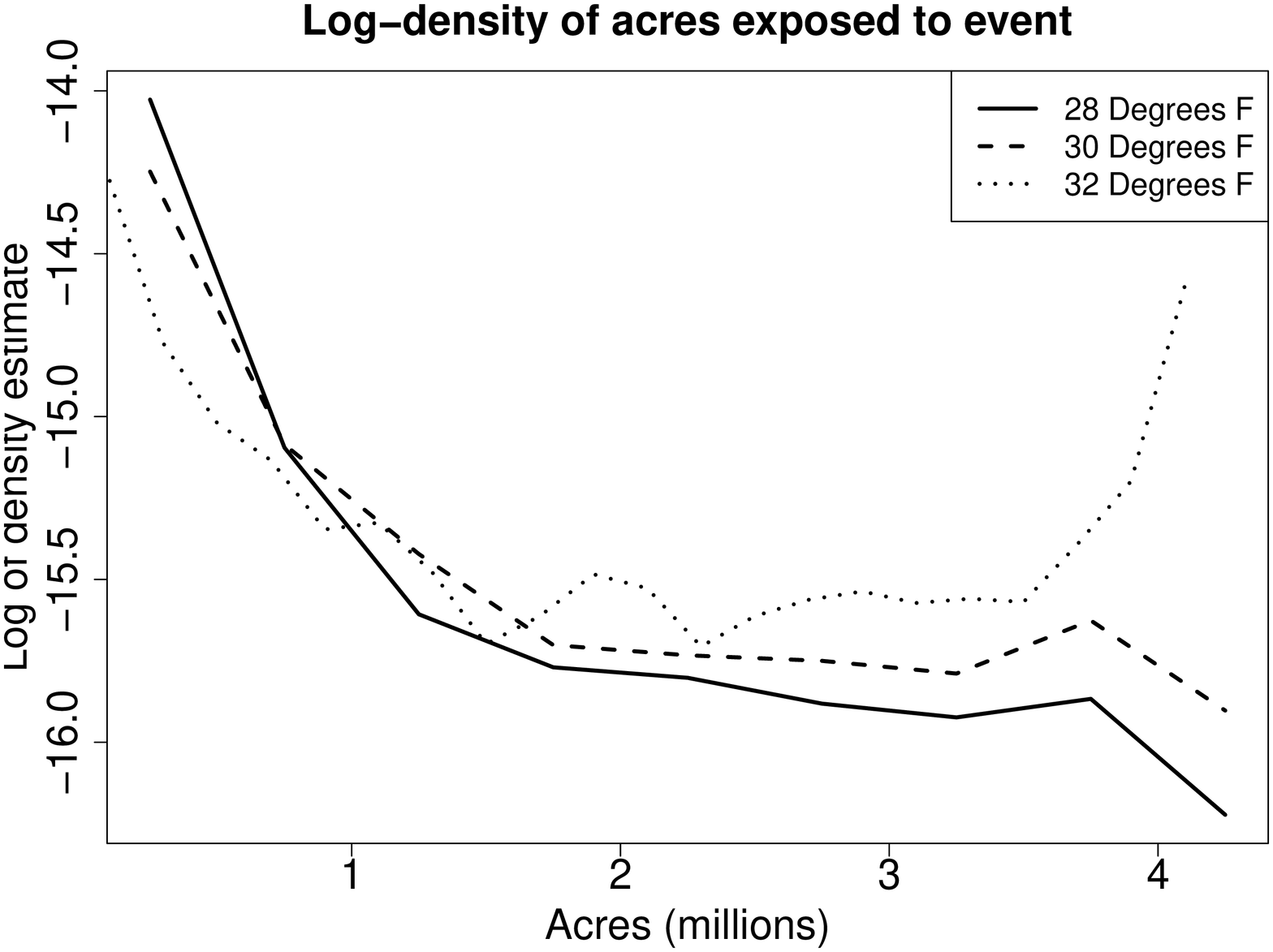}
\caption{A graph showing the number of acres of cotton exposed to minimum temperature event, taken from 10,000 simulations.  The temperature events are minimum October temperature falling below 32 degrees F (dotted line), 30 degrees F (dashed line), and 28 degrees F (solid line).}
\label{fig:count}
\end{center}
\end{figure}

\section{Discussion}
One area which we have not discussed is model selection.  When fitting max-stable processes, there are choices of both the structure of the max-stable model (Schlather vs. alternatives) and of the spatial correlation function within the model.  Ideally, we would like a method of deciding amongst those models.  \citet{varin05} introduced the composite likelihood information criteria (CLIC), which works well with a composite likelihood framework, but this does not have a logical Bayesian interpretation.  The typical approach used in approximate Bayesian computing has been to obtain approximate Bayes factors, which is the more logical choice within the Bayesian context.  Both \citet{pritchard99} and \citet{blum10} followed this approach.  However, recent literature has cast a shadow over the quality of approximating Bayes factors using ABC methods \citep{robert10, sisson10}.  \citet{robert10} in particular found that even with sufficient statistics for two models under consideration, the Bayes factor obtained from ABC cannot always be trusted.  In the more realistic setting with in-sufficient statistics and thus wider loss of information, there is even less reason to trust an ABC Bayes factor.  We are left without a clear ABC model selection procedure at this time, and in the application above, we used the standard CLIC as a first step to determine which model might be best, and then proceeded with ABC.

Approximate Bayesian computing is shown to outperform MCLE in a statistically significant way only for the three short-range max-stable processes.  Still, the results are significant and of value.  We have demonstrated there exists a class of max-stable processes for which an ABC algorithm outperforms the competing MCLE approach.  Furthermore, the simulations in this paper are only exploratory, and in no way exhaust the full range of ABC implementations possible.  There are open questions as to how quartets, quintets, or higher order $k$-tuples may be incorporated into an improved summary statistic, and also there are open questions as to how more efficient ABC samplers could allow the threshold $\epsilon$ to be reduced and thus improve the ABC posterior approximation.  We are continuing to work on these questions, but feel it is a positive development to show an implementation of ABC which outperforms MCLE for short-range processes.  This implementation should serve as a foundation on which improved ABC implementations can be built.

The computational cost of the ABC tripletwise method is appreciably higher than the competing composite likelihood method.  However, for those comfortable with parallel or adaptive computing, the cost is measured in hours, not days or weeks.  Despite this drawback, the ABC approach offers several advantages.  The simulation study has provided sufficient evidence that ABC tripletwise method can result in a lower mean square error when compared to the composite likelihood method.  The method can in principle be extended beyond triplets to $k$-tuples for any $k>3$, and as computational cost falls such implementations should become easier and faster.  And finally, the method very naturally incorporates parameter uncertainty into prediction, which is a central purpose of the field of extremes.

\section*{Acknowledgements}
Helpful feedback and insightful comments from the reviewers and editorial board of \textit{Computational Statistics and Data Analysis} are gratefully acknowledged.


\end{document}